\documentclass[twocolumn,showpacs,preprintnumbers,amsmath,amssymb,prb]{revtex4}
\usepackage{graphicx}
\usepackage{dcolumn}
\usepackage{bm}

\begin{document}

\title{Charge density waves enhance the electronic noise of manganites}

\author{C. Barone$^{1,2}$, A. Galdi$^{1,3}$, N. Lampis$^{1}$, L. Maritato$^{1,2}$, F. Miletto Granozio$^{1}$, S. Pagano$^{1,2}$, P. Perna$^{1}$, M. Radovic$^{1}$,}
\author{U. Scotti di Uccio$^{1}$}
 \altaffiliation{corresponding author} \email{scotti@na.infn.it}

\affiliation{$^{1}$CNR-INFM Coherentia, Complesso Universitario Monte S. Angelo, Napoli, Italy\\
$^{2}$Dipartimento di Matematica e Informatica, Universit\`a  di Salerno, Baronissi (SA), Italy\\
$^{3}$Dipartimento di Fisica "E.R. Caianiello", Universit\`a di Salerno, Baronissi (SA), Italy}

\date{\today}

\begin{abstract}
The transport and noise properties of Pr$_{0.7}$Ca$_{0.3}$MnO$_{3}$ epitaxial thin films in the temperature range
from room temperature to $160~K$ are reported. It is shown that both the broadband 1/f noise properties and the
dependence of resistance on electric field are consistent with the idea of a collective electrical transport, as
in the classical model of sliding charge density waves. On the other hand, the observations cannot be reconciled
with standard models of charge ordering and charge melting. Methodologically, it is proposed to consider
noise-spectra analysis as a unique tool for the identification of the transport mechanism in such highly
correlated systems. On the basis of the results, the electrical transport is envisaged as one of the most
effective ways to understand the nature of the insulating, charge-modulated ground states in manganites.
\end{abstract}

\pacs{72.70.+m, 72.15.Nj, 75.47.Lx}

\maketitle

\section{INTRODUCTION}
The narrow-band, mixed-valence manganites are since over ten years under the focus of a wide scientific community,
mainly because of the "colossal" responses of their electronic properties to magnetic field and to other external
perturbations. It was soon recognized that many effects are related to charge-modulated phases, described in the
earlier models as due to the full charge disproportionation on the Mn site, and to the consequent crystallization
(i.e. "charge ordering", CO) of a lattice of ordered Mn$^{3+}$/Mn$^{4+}$ ions. Such ordered systems can be
abruptly driven to the conducting state by several kinds of perturbation, including electrical \cite{Asamitsu} and
magnetic field \cite{Tomioka}, optical \cite{Rini}$^{,}$\cite{Fiebig}, x ray radiation \cite{Kiryukhin}, and
pressure \cite{Hwang}. Within the classical interpretation, such effects are ascribed to the "melting" of the CO
phase \cite{Asamitsu}$^{,}$\cite{Tomioka}$^{,}$\cite{Fiebig}, meaning that the perturbation destroys the ordered
state and frees the charge carriers, enhancing the conductivity up to several orders of magnitude. Recently, this
interpretation has been challenged by several authors, on the basis of an increasing body of evidence pointing to more
complex descriptions of both the static and the dynamic properties of such compounds. As we discuss in the
following, the results of our research on the transport properties of the Pr$_{0.7}$Ca$_{0.3}$MnO$_{3}$ manganite
also support the same view.

Pr$_{1-x}$Ca$_{x}$MnO$_{3}$ (PCMO) is a very interesting system, where charge, orbital, lattice, spin degrees of
freedom strongly interact and determine the macroscopic properties. The phase diagram of bulk PCMO \cite{Raveau}
includes an insulating ferromagnetic (FM) phase at $x < 0.3$, where the CO state is suppressed. In the higher
doping range ($0.3 < x < 0.75$), an ordered phase is found below the transition temperature $T_{CO}$. $T_{CO}$
exceeds $200~K$ in the range $0.3 < x < 0.5$, then it reaches its maximum at the "commensurate" stoichiometry $x =
0.5$, and it finally drops at higher doping \cite{Pollert}$^{,}$\cite{Jirak}. The structure of PCMO at $0.3 < x <
0.5$ has been recently subject of extensive investigation \cite{Daoud}$^{,}$\cite{Jooss1}, in order to settle the
issue of charge localization. Both quoted papers indicate that the lattice hosts Zener polarons, i.e. dimers of Mn
ions resonantly sharing an electron. The ordered state is then depicted as a Zener polaron lattice, that is well
ordered at $x = 0.5$ and that is stable down to $x = 0.3$, where the extra Mn$^{3+}$ ions are hosted as
impurities. The conduction mechanism in PCMO above the ordering transition is well described by models based on
the motion of thermally activated Jahn Teller polarons \cite{Jooss2}. The same authors show that, below the
ordering transition temperature, the conduction is instead due to a cooperative mechanism involving the coherent
motion of clusters of Zener polarons. Above a certain threshold current, a transition toward a light-polarons
state sets in, determining the giant electroresistive effect.

The existence of a cooperative mechanism of motion is also found in the classical model of charge density waves
(CDW), introduced in his pioneering work by Peierls \cite{Peierls} and later by Gr\"{u}ner \cite{Gruner}. Within
this model, the 1-dimensional electron density is (incommensurately) modulated, due to the interaction with
phonons. The CDW picture has been invoked in recent papers on manganites. S. Cox \textit{et al.}
\cite{Cox}$^{,}$\cite{Cox1} base this claim on an accurate characterization performed by resorting
to transmission electron microscopy and to transport measurement on La$_{0.5}$Ca$_{0.5}$MnO$_{3}$ and
Pr$_{0.7}$Ca$_{0.3}$MnO$_{3}$ films. Nucara \textit{et al.} \cite{Nucara} provided support to the same ideas on
the base of optical conductivity measurements on several compositions of manganites, showing spectral features
that are associated with the typical excitations (phasons and amplitudons) of a CDW system. Previous data
supporting this concept had also been reported, based on nonlinear transport \cite{Wahl}, optical spectroscopy
\cite{Kida}$^{,}$\cite{Calvani}, and specific heat measurements \cite{Cox2}.

The debate is still quite hot, since it is also argued \cite{Schmidt} that the CDW regions of the phase diagrams,
if any, are limited to restricted areas at the border with a metallic phase. Different intermediate situations may
probably take place in different materials, according to bandwidth, doping, quenched disorder and strain. Under
this point of view, PCMO with a doping level $x < 0.5$ is an interesting subject of investigation. Furthermore, a
clear-cut distinction between the CDW slippage (that is a coherent motion) and the disordered creep of a rigid CO
lattice is not straightforward. The experimental identification of a CDW state by only dc transport measurements
is quite cumbersome and, probably, it cannot be unambiguous. In this paper, concerning the electric properties of
high quality Pr$_{0.7}$Ca$_{0.3}$MnO$_{3}$ films, we propose instead that the transport regime is better
identified by investigation of the electronic noise as a function of the temperature and of the applied electric
field.

\section{EXPERIMENTAL}
The samples adopted in this work are thin ($10~nm$) Pr$_{0.7}$Ca$_{0.3}$MnO$_{3}$ (PCMO) epitaxial films grown
under tensile strain on (100) SrTiO$_{3}$ (STO) substrates. Details of the complex PLD chamber where the growth took place are reported in
Ref.~\onlinecite{Radovic}. Briefly, the films are grown in a $0.1~mbar$ oxygen atmosphere, at a rate of 2 unit cells per minute, by resorting to a Kr-F excimer laser ($248~nm$) that radiates the target with $50~mJ$/$2.4~mm^{2}$ effective fluency.
\begin{figure}[!t]
\resizebox{1.0\columnwidth}{!}{%
  \includegraphics{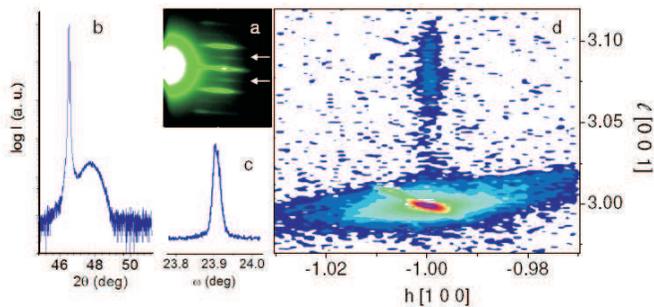}
} \caption{(color online) (a) RHEED patterns of a PCMO film; the arrows indicate the (110) and reflections. (b)
$\theta /2 \theta$ showing the (002) STO and the (004) PCMO peaks; (c) the (004) PCMO rocking curve; (d) RSM
showing the (310) STO and the (622) PCMO.} \label{fig:Structural-Fig}
\end{figure}
The RHEED patterns collected during deposition [Fig.~\ref{fig:Structural-Fig}(a)] show flat surfaces (2-dimensional growth) preserved until the end of the deposition. The bulk Pr$_{0.7}$Ca$_{0.3}$MnO$_{3}$ is orthorhombic, with lattice parameters $a=0.5426~nm$, $b=0.5478~nm$,
$c=0.7679~nm$. The substrate lattice spacing, $a_{s}=0.3905~nm$, is matched to Pr$_{0.7}$Ca$_{0.3}$MnO$_{3}$,
being $c \approx 2a_{s}$ and $a \approx b \approx \sqrt{2}a_{s}$. Such a pseudocubic orthorhombic film can be
accommodated on STO in several ways \cite{Ricci}$^{,}$\cite{Awaya}. Our x-ray diffraction characterization is
indicative of a c-axis orientated single domain with vertical spacing $c=0.7622~nm$,
Fig.~\ref{fig:Structural-Fig}(b). The in-plane Pr$_{0.7}$Ca$_{0.3}$MnO$_{3}$ lattice is fully strained
[Fig.~\ref{fig:Structural-Fig}(d)], with the a and b axes aligned to the (110) STO, so that the unit cell is
tetragonal. Such epitaxial accommodation results in a relatively small strain ($\epsilon_{a}=0.018$; $\epsilon _{b}=0.008$;
$\epsilon _{c}=-0.007$) and a high crystal quality [Fig.~\ref{fig:Structural-Fig}(c)].

The stress applied by the substrate, in principle, may set a difference between film and bulk properties. The bulk samples, with $x = 0.3$, are just on the borderline, in the bulk phase diagram, between the insulating ferromagnetic and the insulating antiferromagnetic, charge ordered, phase. The films have an intermediate behavior, that will be thoroughly discussed elsewhere and that we briefly anticipate here. The samples are ferromagnetic below $120~K$, even though the large difference between zero-field cooled and field cooled magnetization vs. temperature, the high coercivity and other experimental signatures suggest a glassy ferromagnetic state. On the other hand, as we show and discuss in the following, the samples also undergo a transition that we interpret as due to charge modulation at about $230~K$. In our understanding, this charge-modulated state is stable in the range $120~K$ - $230~K$, while it is melted at about $230~K$; below $120~K$ there is indication of coexistence of charge and ferromagnetic ordering.

\begin{figure}[!t]
\resizebox{1.0\columnwidth}{!}{%
  \includegraphics{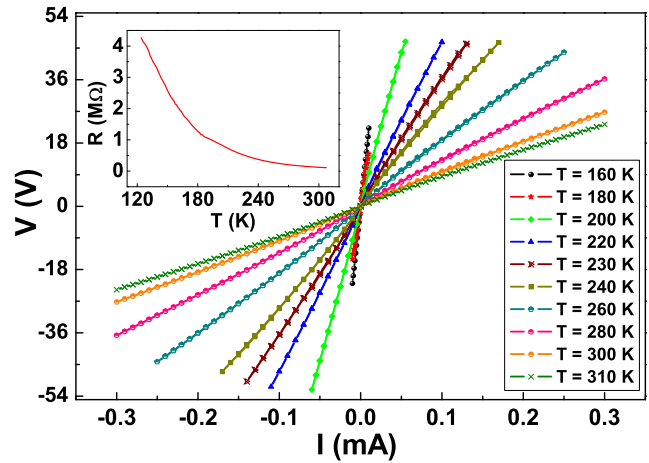}
} \caption{(color online) I-V curves in range $160~K$$<$T$<$$310~K$. Inset: R vs. T for the same sample.}
\label{fig:dcTransport-Fig}
\end{figure}

The transport measurements were carried out in the standard four-probe configuration with in-line geometry. The
electrical contacts were obtained by bonding gold wires on silver stripes, deposited at $2~mm$ one from each
other, so that the lines of current were always at $45^{\circ}$ with the a axis. The superstructure in the charge
ordered phase, on the other hand, is known to modulate the bulk properties along the a axis. We expect this also
holds for our films, due to the low strain of the structure. Only a few reports regard structural investigations
of charge ordering in films; a modulation along the a axis is reported in the similar compound
La$_{0.5}$Ca$_{0.5}$MnO$_{3}$ films \cite{Cox}. The resistance vs. temperature R(T)
(Fig.~\ref{fig:dcTransport-Fig}) was measured in the current-pulsed mode, by biasing the samples with an active
current source. No evident feature is present at the expected charge modulation temperature as reported in bulk
($T_{CM} \approx 230~K$), while a small feature is found at $180~K$, corresponding to the Neel temperature $T_{N}$
\cite{Pollert}. A signature of the charge modulation taking place below $230~K$ is provided by the temperature
dependent I-V characteristics reported in the same picture. The data were collected by resorting to $1~ms$ current
pulses to avoid Joule heating. Below $230~K$, a slight non-linearity sets in. No hysteretic features were
observed, that we attribute to the limited value of the applied electric field.

\begin{figure}[!t]
\resizebox{1.0\columnwidth}{!}{%
  \includegraphics{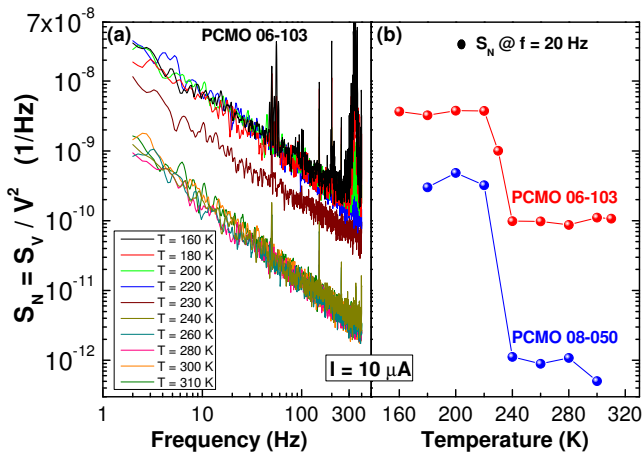}
} \caption{(color online) (a) Normalized voltage spectral density $S_{N}$ at fixed current $I=10~\mu A$ between
$160$ and $310~K$. The peaks superimposed to the 1/f component are due to external noise sources. (b) $S_{N}$ at
$f=20~Hz$ for two distinct samples.} \label{fig:SNvsT-Fig}
\end{figure}

Voltage noise spectral density measurements were carried out both in the four and in the two-probe configurations.
A major effort was dedicated to minimizing the electrical noise generated by the contacts. The problem was
circumvented by resorting to an improved measurement technique, described in Ref.~\onlinecite{Barone-NoiseRed},
aimed to eliminate artifacts due to the contact noise contribution. The normalized voltage spectral density
$S_{N}(f)=S_{V}(f)/V^{2}$ reveals a unique 1/f component between $160$ and $310~K$ [Fig.~\ref{fig:SNvsT-Fig}(a)].
The most striking feature [Fig.~\ref{fig:SNvsT-Fig}(b)] is a step-like increase of the $S_{N}$ amplitude by
one-two orders of magnitude, at about $230~K$ \cite{Mercone}. Several performed checks exclude that such unusual
behavior is an artifact. Figure~\ref{fig:SNvsT-Fig}(b) shows for example the comparison of data regarding a fresh
sample (PCMO 08-050), and a two years old one (PCMO 06-103), the latter showing a much higher noise intensity,
probably due to development of structural defects or to the interaction with air. In spite of the overall
different quantitative behavior, the two samples share the essential feature which is under investigation in this
work.

\begin{figure}[!t]
\resizebox{1.0\columnwidth}{!}{%
  \includegraphics{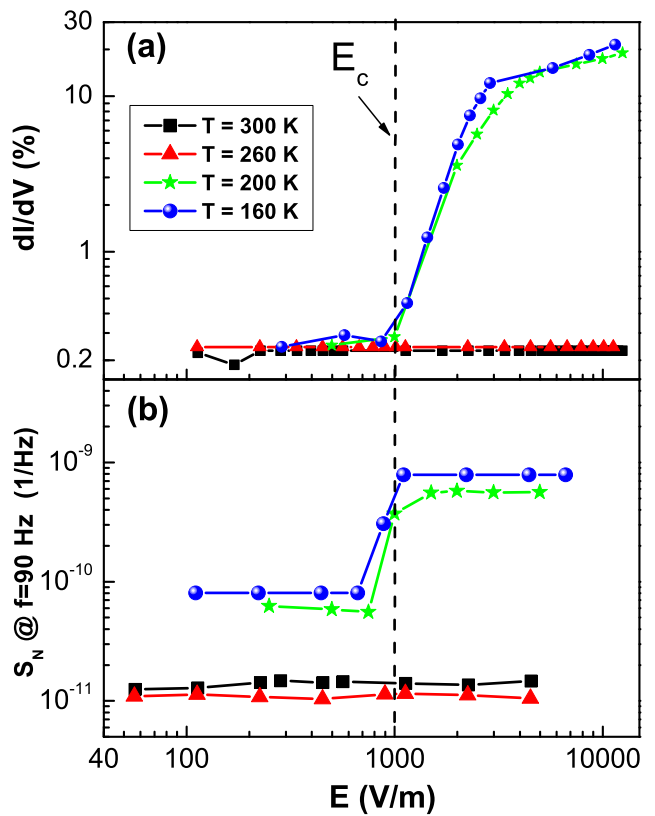}
} \caption{(color online) Electric field dependence of the normalized differential conductivity (a) and of the
normalized spectral density at a frequency $f=90~Hz$ (b).} \label{fig:ElectricField-Fig}
\end{figure}

In order to get a further insight into the physical properties of Pr$_{0.7}$Ca$_{0.3}$MnO$_{3}$ below $230~K$, we
compared the electric field dependence of the differential conductivity and of the normalized spectral density
(Fig.~\ref{fig:ElectricField-Fig}). Above $230~K$, the normalized differential conductivity [defined as
$\left(dI/dV\right)/ \sigma_{0}$, where $\sigma_{0}$ is the conductivity at zero voltage] is constant vs. the
electric field $E$, that is, the sample exhibits an Ohmic behavior. In this regime, the normalized voltage
spectral density is independent on $E$, as well. In the low temperature phase, on the contrary, both plots are
characterized by a threshold $E_{c} \simeq 1000~V/m$ at which a further very strong step-like increase of the
$S_{N}$ amplitude is found.

\section{DISCUSSION}
The absence of any sign on the R(T) curve and the step-like noise increase are not easy to reconcile with a
"strong coupling" CO picture in which the ordered network of charges is tightly localized on Mn sites. In such a
case we would rather expect i) a step increase of resistivity and ii) a noise reduction when the "charge crystal"
condenses from the "liquid" phase. Phenomenologically, the CDW model can instead explain these findings:

i) the continuity of R at $T_{CM}$ can be understood in the framework of a phenomenological two-fluid model (we
remind, to this aim, the noticeable, formal similarities between the CDW state and the superconducting state). In
such model, the relative weight of the ordered phase (that in the classical view is not spatially separated)
starts from $0$ at the thermodynamic transition temperature and it approaches $1$ only at $T = 0$. In the
microscopic description of the CDW state, the same gradual effect on resistivity can be envisaged as due to the
temperature dependence of the thermal excitations above the gap. The low temperature value 2$\Delta$(T=0) in the
excitation spectrum of the CDW state has been estimated $\approx$ 100-200 meV in Ref.~\onlinecite {Cox1}. We note that
the quoted mechanisms are related to dimensionality issues, namely, they may have different impact in bulk
and in thin films. Actually, PCMO bulk samples generally show sharper signatures of the ordering transition in
R(T) plots with respect to thin films.

ii) Intuitively, the transition to an ordered state should reduce the electronic noise. Even if the conduction
mechanism is due to creep of charges at the border of the "charge-crystal" region, one may imagine that the noise
is at most unchanged, since the creep is not qualitatively different from the polaron hopping characterizing the
melt state. On the contrary, the presence of an excess broadband 1/f noise is a well established feature of CDW
systems \cite{Zettl}$^{,}$\cite{Bhatta}. Such 1/f noise stems from the many metastable states of the CD
condensate, that may be seen as long-wavelength phase modulations of the CDW order parameter. Each such state
suffers from different pinning forces due to the lattice defects and it is therefore characterized by a different
electrical resistance. At a given value of current flow, the thermal activation of transitions between such states
determines a resistance fluctuation that results in the extra voltage noise \cite{Cava}.

Once the CDW model is accepted, more details that connect the noise spectra to the dc transport properties are
better understood. The onset of nonlinearity above a critical field (in the I-V characteristics below $T_{CM}$) is
also a typical signature of CDW systems. In the Lee-Rice model \cite{Lee} $E_{c}$ is the depinning field, above
which the CDW can slide. Our value of $1000~V/m$ compares well with other reported data
\cite{Cox}$^{,}$\cite{McCarten}. More importantly, we observed that the enhancement of noise starts just below the
threshold, and it is completed at $E_{c}$, that is, the same scale of fields determines both the nonlinearity and
the noise enhancement. This is just what one expects in a CDW state \cite{Bhatta}.

A further check is based on the comparison between our data and the theoretical expectations for the behavior of
1/f noise. According to Ref.~\onlinecite{Bhatta}, assuming that the pinning force fluctuations are the source of
noise in the chordal resistance $R=V/I$, the voltage noise is given by: $\left< \delta V^{2}\right> = I^{2}\left(
\partial R / \partial V \right)^{2} \left< \delta V_{c}^{2}\right>$, where $V_{c}$ is the threshold voltage
associated with $E_{c}$.
\begin{figure}[!t]
\resizebox{1.0\columnwidth}{!}{%
  \includegraphics{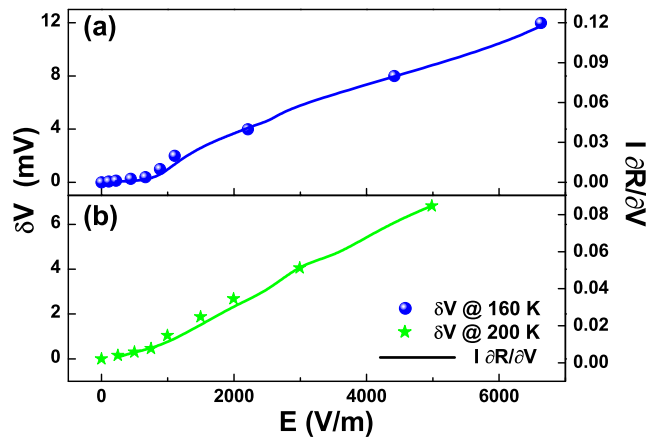}
} \caption{(color online) $\delta V$ (points) and $I \partial R / \partial V$ (solid lines) vs. $E$, at $160~K$
(a) and $200~K$ (b). Right scales have been adjusted for best fitting with noise data.} \label{fig:PRLcomp-Fig}
\end{figure}
This expression was employed in Ref.~\onlinecite{Bhatta} to establish a quantitative relation between the
non-linear conductance and the noise level. By applying the same analysis to our data below $T_{CM} \approx
230~K$, a quantitative agreement is found. In Fig.~\ref{fig:PRLcomp-Fig}(a) and (b), the average voltage noise
$\delta V=\sqrt{\left< \delta V^{2}\right>}$, where $\left< \delta V^{2}\right>$=$\int_{1Hz}^{100kHz}dfS_{V}(f)$,
is plotted vs. the electric field $E$; the behavior is compared to the values of of $I \partial R / \partial V$
vs. $E$, as determined by the corresponding I-V characteristic of the sample. The data fairly overlap. In the
Ohmic regime above $T_{CM}$, on the other hand, no such correlation is possible, because $I \partial R / \partial
V \simeq 0$ while $\delta V$ grows linearly with the electric field [see, e.g.,
Fig.~\ref{fig:ElectricField-Fig}(b), where $S_{N} = const$ at $260~K$ and $300~K$ implying $\delta V \propto E$).
Furthermore, we can evaluate the coherence volume $\lambda^3$, that is the typical volume over which the CDW phase
is deformed in going from one metastable state to another, by the equation
\begin{equation}
S_{V}\left(f,T\right) = I^{2}\left\{S_{R}\left(f,T\right)+\left[\frac{\partial R}{\partial V}\right]^{2}E_{c}^{2}
\lambda ^{3}\left(T\right) \frac{l}{A} S\left(f,T\right)\right\} \label{eq:CDWnoise}
\end{equation}
that we adapted from Ref.~\onlinecite{Bhatta} to include the fluctuations in the normal state. Here $l$ and $A$
are sample length and area, $S_R\left(f,T\right)$ is the spectral density of resistance fluctuations and
$S\left(f,T\right)$ is the normalized spectral weight function of the transition rate between the metastable
states of the condensate. By integrating Eq.~(\ref{eq:CDWnoise}) over the investigated range of frequency ($1~Hz$
- $100~kHz$) and by neglecting the fluctuation term related with $S_R$, which is more than one order of magnitude
lower than the measured noise, $\lambda$ is determined in terms of the effective voltage noise $\left< \delta
V\right>$. By using our estimation $E_{c}=1000~V/m$, we get $\lambda \simeq 400~nm$ at $T=160~K$. This is close to
the value of the Lee-Rice coherence length $\lambda_{LR} \simeq 300~nm$, based on the relation of
Ref.~\onlinecite{Lee}, for $E_{F} \simeq 1~eV$ and $Q \simeq 10^{10}~m^{-1}$ (see, e.g., Ref.~\onlinecite{Cox} for
La$_{0.5}$Ca$_{0.5}$MnO$_{3}$). Moreover, we find $\lambda \simeq 100~nm$ at $T=200 K$. The decrease of $\lambda$,
when the temperature approaches $T_{CM}$, is consistent with the Lee-Rice result $\lambda _{LR} \rightarrow 0$ as
$T \rightarrow T_{CM}$, when assuming that $\lambda$ scales with$\lambda _{LR}$.

Finally, we would like to mention that other physical mechanisms, that may be invoked as source of the extra noise
in the charge modulated phase, don't lead to the observed broadband 1/f behavior. As an instance, the tunneling of
charges through the grain boundaries between separate regions (may take place in the case of phase separation)
results in either a Lorentzian spectral density, or a white spectrum \cite{Savo}. The random telegraph noise, that
also may play a role in phase-separated compounds, is related to the occurrence of strongly nonlinear electrical
transport and to a Lorentzian spectral density in the very low frequency region
\cite{Barone-LSMO}$^{,}$\cite{Bid}, at odds with our findings of weak nonlinearity and pure 1/f behavior.

\section{CONCLUSIONS}
In conclusion, we investigated the nature of the charge modulated state in Pr$_{0.7}$Ca$_{0.3}$MnO$_{3}$ thin
films by measuring the properties of its current-carrying excited states. The principal conclusion of this work is
that the voltage noise measurements proved to be a unique probe of the transport mechanism that is activated below
the ordering transition temperature. The observed step-like increase of the broadband 1/f noise level at lowering
the temperature is peculiar and calls for a specific interpretation. In most conventional materials, the 1/f
component is in fact either temperature independent, or slowly decreasing at decreasing temperature.

Our results are in fact consistent with the expectations based on the CDW model, formerly adopted for
1-dimensional metals and recently extended to the description of narrow band manganites. Qualitatively, it is hard
to reconcile the observation of the 1/f extra noise with the picture of a rigid lattice of charges
crystallizing from a melt phase of polarons, where a noise reduction may well be expected. Moreover, it is still
more difficult to understand why the subsequent application of electric fields above some threshold (that in the
CO scenario should contribute to melt the lattice again, coming back to the pristine conditions) should further
increase the noise.

We also could verify that the extra noise mechanism is connected with the onset of nonlinearity in the I-V
characteristic of samples. Actually, the detailed description of the voltage noise features fully agree with
current models of voltage noise generation in the CDW state.

This is, in our opinion, a significant result, in view of the still open debate on the CDW state in manganites. The narrow-band oxides are such a complex system, highly sensitive to doping, stress, etc., that more data on the issue of the CDW state and of its implications are actually welcome, to assess under what circumstances and in what samples the model is applicable. We stress, however, that the model itself may need some revision in its theoretical foundations, to account for the specific features of perovskitic, narrow band oxides, as compared to the simple, 1-dimensional metals that are considered in the original work of Peierls.

The implications of the CDW model are of great interest in connection with the fundamental properties of the
charge modulated state, such as its commensurability with the lattice. In this respect, we observe that the
investigated PCMO samples, with $x = 0.3$ doping level, are far from the ideally "commensurate" $x = 0.5$ value.
The consistent applicability of the CDW model to this case, in connection with other published results on
different doping levels (and materials), is an indication of the robustness of such state, meaning that it seems
to extend its range of application on a wide portion of the phase diagram.

The emerging picture suggests that in manganites the charges may be displaced with more easiness then formerly
believed, thus envisaging a connection with the CDW dynamics, that requires a deformable charge lattice. Such a
finding is relevant to many aspects of the physics of manganites, and, in general, of correlated oxides. While in
a CO state a phase separation picture between localized CO regions and ferromagnetic delocalized regions is
expected, novel "electronically soft" phases \cite{Milward}, where the two order parameters (charge modulation and
magnetization) coexist within the same domain, have been proposed to be compatible with the CDW case. This is
pertinent to the physics of our samples, where both transitions are found. Together with recent claims of sliding
charge density waves in cuprates \cite{Blumberg}, these results suggest that the CDW mechanism should be
introduced as one of the bricks necessary to build up our comprehension of strongly correlated transition metal
oxides. Such interpretation also allows to bridge the established approach that envisages CDW compounds as chaotic
systems to the recent claims that the understanding of correlated oxide properties should be pursued in the
broader context of complexity \cite{Dagotto}.

\begin{acknowledgments}
The authors wish to thank Vittorio Cataudella for the fruitful discussions and for his help in the theoretical
assessment of the work.
\end{acknowledgments}

\end{document}